\documentclass[twocolumn,english,prl,superscriptaddress,showpacs,longbibliography,fixfloat,notitlepage]{revtex4-1}
\usepackage[T1]{fontenc}
\usepackage[latin9]{inputenc}
\usepackage{color}
\usepackage{amstext}
\usepackage{amssymb}
\usepackage{graphicx}

\makeatletter
\usepackage[colorlinks,citecolor=darkblue,linkcolor=darkred,urlcolor=darkblue] {hyperref}
\usepackage{xcolor}
\definecolor{darkblue}{rgb}{0.1,0.2,0.6} 
\definecolor{darkred}{rgb}{0.8,0.1,0.2}
\renewcommand{\BibitemShut}[1]{}

\makeatother

\usepackage{babel}
\begin{document}
\global\long\def\E{\mathrm{e}}
\global\long\def\D{\mathrm{d}}
\global\long\def\I{\mathrm{i}}
\global\long\def\mat#1{\mathsf{#1}}
\global\long\def\vec#1{\mathsf{#1}}
\global\long\def\cf{\textit{cf.}}
\global\long\def\ie{\textit{i.e.}}
\global\long\def\eg{\textit{e.g.}}
\global\long\def\vs{\textit{vs.}}
 \global\long\def\ket#1{\left|#1\right\rangle }
\global\long\def\etal{\textit{et al.}}
\global\long\def\tr{\text{Tr}\,}
 \global\long\def\im{\text{Im}\,}
 \global\long\def\re{\text{Re}\,}
 \global\long\def\bra#1{\left\langle #1\right|}
 \global\long\def\braket#1#2{\left.\left\langle #1\right|#2\right\rangle }
 \global\long\def\obracket#1#2#3{\left\langle #1\right|#2\left|#3\right\rangle }
 \global\long\def\proj#1#2{\left.\left.\left|#1\right\rangle \right\langle #2\right|}

\title{Time-dependent variational principle in matrix-product state manifolds:
pitfalls and potential}

\author{Benedikt Kloss }

\affiliation{Department of Chemistry, Columbia University, 3000 Broadway, New
York, New York 10027, USA}
\email{bk2576@columbia.edu}

\author{Yevgeny Bar Lev }

\affiliation{Department of Chemistry, Columbia University, 3000 Broadway, New
York, New York 10027, USA}

\affiliation{Department of Condensed Matter Physics, Weizmann Institute of Science,
Rehovot 76100, Israel}

\affiliation{Max-Planck-Institut für Physik komplexer Systeme, 01187 Dresden,
Germany}
\email{yevgeny.barlev@weizmann.ac.il}

\author{David Reichman}

\affiliation{Department of Chemistry, Columbia University, 3000 Broadway, New
York, New York 10027, USA}
\begin{abstract}
We study the applicability of the time-dependent variational principle
in matrix product state manifolds for the long time description of
quantum interacting systems. By studying integrable and nonintegrable
systems for which the long time dynamics are known we demonstrate
that convergence of long time observables is subtle and needs to be
examined carefully. Remarkably, for the disordered nonintegrable system
we consider the long time dynamics are in good agreement with the
rigorously obtained short time behavior and with previous obtained
numerically exact results, suggesting that at least in this case the
apparent convergence of this approach is reliable. Our study indicates
that while great care must be exercised in establishing the convergence
of the method, it may still be asymptotically accurate for a class
of disordered nonintegrable quantum systems.
\end{abstract}
\maketitle
\emph{Introduction}.\textemdash{} The numerically exact simulation
of the dynamics of strongly interacting quantum systems is a grand
challenge in condensed matter science. For ground states of gapped
one-dimensional systems with short-range interactions, the density
matrix renormalization group (DMRG) proves to be a powerful and efficient
approach \cite{White1992,Schollwock2011}. Its success is linked to
the fact that the ground states of these systems are optimally representable
by matrix product states (MPS), with a moderate number of variational
parameters, normally referred to as the bond dimension. While DMRG
has been extended into the time-domain, the timescales that may be
reached are usually quite short as a consequence of correlations that
develop within the propagated wavefunction \cite{White1992,Schollwock2011}.
Time evolution tends to quickly displace states from the space efficiently
representable by MPS, leading to a rapid (typically exponential) growth
of the bond dimension. If the bond dimension of the wavefunction is
not dynamically adjusted to accommodate the growing correlations in
the wavefunction the dynamics quickly becomes approximate and nonunitary.
It is possible to construct a unitary time-propagation scheme on the
manifold of MPS with a fixed bond dimension using the Dirac-Frenkel
time-dependent variational principle (TDVP) \cite{PSP:2040328,frenkel1934mechanics,Haegeman2011,Haegeman2016}.
This principle, which is rather generic, projects an infinitesimal
time evolution under the Hamiltonian to a variational manifold which
the resulting wavefunction is restricted to occupy. An advantage over
conventional DMRG techniques is that the TDVP can be applied to a
more general class of states, such as tree tensor network states,
thus potentially opening the door to efficiently simulating higher
dimensional systems as well as systems with long-ranged interactions.

The description of transport properties requires the investigation
of large system sizes and long times, a limit which is sometimes referred
to as the hydrodynamic limit. While this limit appears to be out of
reach for numerically exact methods, an approximate coarse-grained
treatment might be sufficient to obtain accurate macroscopic observables
like transport coefficients, analogous to the success of classical
hydrodynamics. In this respect the TDVP is particularly attractive,
since it generates effectively chaotic classical dynamics in the space
of variational parameters which obey a set of macroscopic conservation
laws, such as those associated with the total number of particles
and the total energy \cite{Haegeman2013}. Indeed, a surprisingly
fast convergence of the heat diffusion constant with respect to bond
dimension has been very recently reported for a nonintegrable spin
chain \cite{Leviatan2017}.

In this work, we examine the applicability of TDVP for the long time
description of quantum interacting systems. While the method cannot
be expected to work for quantum integrable systems (c.f. generalization
of hydrodynamic approaches to such systems \cite{PhysRevLett.117.207201}),
by utilizing the exact solvability of such systems we show that the
long time limit, which is necessary to obtain hydrodynamic observables,
and the large bond-dimension limit, where the method becomes numerically
exact, do \emph{not }generically ``commute.'' In particular, the
apparent convergence of hydrodynamic observables with the bond dimension
does \emph{not} guarantee the accuracy of the result, which has to
be established by other means. This problem persists also for nonintegrable
systems, although for the case of a disordered \emph{nonintegrable}
quantum system that we consider, this problem appears to be ameliorated.

\emph{Theory}.\textemdash{} The Hilbert-space dimension of a quantum
lattice systems scales exponentially with the size of the system.
Any wavefunction in the Hilbert space can be written as a matrix product
state (MPS),

\begin{equation}
\ket{\Psi[A]}=\sum_{\{s_{n}\}=1}^{d}A^{s_{1}}(1)A^{s_{2}}(2)\dots A^{s_{N}}(N)\ket{s_{1}s_{2}\dots s_{N}}\label{eq:MPS-gen}
\end{equation}
where $d$ is the local Hilbert space dimension, $A^{s_{i}}(i)\in\mathbb{C}^{D_{i-1}\times D_{i}}$
are complex matrices and $D_{0}=D_{N}=1$, such that the product of
matrices evaluates to a scalar coefficient for a given configuration
$\ket{s_{1}s_{2}\dots s_{n}}$. To be an \emph{exact} representation
of the wavefunction the dimension of the matrices the bond dimension
must scale exponentially with the systems size. Typically one approximates
the wavefunction by truncating the dimension of the matrices to a
predetermined dimension with computationally tractable number of parameters.
Exact results are obtained when the approximate dynamics are converged
with respect to the bond dimension.

The time-dependent variational principle (TDVP) allows one to obtain
a locally optimal (in time) evolution of the wavefunction on the manifold
of MPS, $\mathcal{M}_{\mathbf{r}},$ with some fixed bond dimension
$\mathbf{r}$. It amounts to solving a tangent-space projected Schrödinger
equation \cite{Haegeman2016}:
\begin{equation}
\frac{d\ket{\Psi[A]}}{dt}=-\I P_{\mathcal{M}}\hat{H}\ket{\Psi[A]},\label{eq:TDVP}
\end{equation}
where $P_{\mathcal{M}}$ is the tangent space projector to the manifold
$\mathcal{M}_{\mathbf{r}}$.\textcolor{magenta}{{} }Equation (\ref{eq:TDVP})
is solved using a Trotter-Suzuki decomposition of the projector(see
Ref.~\cite{Haegeman2016} for details).

The dynamics generated by the TDVP can be viewed as resulting from
a classical, non-quadratic Lagrangian in the space of variational
parameters \cite{Haegeman2011,Leviatan2017}. It can be shown that
any conserved quantity of the Hamiltonian will be also conserved by
TDVP if the corresponding symmetry group members of the associated
quantity applied to a state in the manifold $\mathcal{M}_{\mathbf{r}}$
do not take it out of the manifold \cite{Haegeman2013}. The nonlinearity
of the equations of motion of TDVP disappears in the limit of infinite
bond dimension, since in this limit the action of the Hamiltonian
on the state keeps it on the manifold for all times. 

\emph{Results}.\textemdash{} We study transport properties of the
one-dimensional XXZ model,
\begin{equation}
\hat{H}=J_{xy}\sum_{i=1}^{N-1}\left(\hat{S}_{i}^{x}\hat{S}_{i+1}^{x}+\hat{S}_{i}^{y}\hat{S}_{i+1}^{y}\right)+\Delta\sum_{i}\hat{S}_{i}^{z}\hat{S}_{i+1}^{z}+\sum_{i=1}^{N}h_{i}\hat{S}_{i}^{z},\label{eq:XXZ-Ham}
\end{equation}
where $h_{i}$ is uniformly distributed in the interval $\left[-W,W\right]$
and $\hat{S}_{i}^{\left(x,y,z\right)}$ are the appropriate projections
of the spin operators on site $i$. In the following, we use $J_{xy}=1$,
which sets the time unit of the problem. Using the Jordan-Wigner transformation
the XXZ model can be mapped to a model of spinless fermions \cite{Jordan1928}.
For $\Delta=0$, the corresponding model is noninteracting and can
be solved exactly. In particular for $W\neq0$ the system becomes
Anderson localized \cite{Anderson1958b}. For $\Delta\neq0$ and at
sufficiently high disorder the system becomes many-body localized
and exhibits a dynamical phase transition \cite{Basko2006a,Abanin2017}
which, for $\Delta=1$, occurs at $W\approx3.5$ \cite{Berkelbach2010a,Luitz2015}. 

To study the dynamical properties of this model in its various limits
we calculate the spreading of a spin-excitation as a function of time,
\begin{equation}
\sigma^{2}(t)=\sum_{i=1}^{L}(\frac{L}{2}-i)^{2}\left\langle \hat{S}_{i}^{z}(t)\hat{S}_{L/2}^{z}(0)\right\rangle .\label{eq:MSD}
\end{equation}
Here the expectation value is calculated at infinite temperature,
namely $\left\langle \hat{O}\right\rangle =\tr\hat{O}/\mathcal{N}$
where $\mathcal{N}$ is the Hilbert space dimension. The spread of
the excitation is analogous to the classical mean-square displacement
(MSD). Transport is characterized by assuming a power law scaling
of the MSD, $\sigma^{2}(t)\sim t^{\alpha}$. For example, a dynamical
exponent of $\alpha=2$ ($\alpha=1$) indicates ballistic (diffusive)
transport. A dynamical exponent $0<\alpha<1$ corresponds to subdiffusive
transport, and $\alpha=0$ for localized systems. We also define a
time-dependent diffusion constant $D(t)$ as the time-derivative of
$\sigma^{2}(t)$ \cite{Steinigeweg2009a,Yan2015,Steinigeweg2017,Luitz2016c}.
Throughout this work the hydrodynamic variable that we will consider
will be the asymptotic spin diffusion coefficient, $\lim_{t\to\infty}D\left(t\right)\to D$.

To calculate the MSD we numerically evaluate the correlation function
starting from a random configuration of up and down spins and also
a random configuration of the disordered field, when appropriate.
By sampling simultaneously both spin configurations and disorder configurations
we obtain the required infinite temperature initial conditions and
disorder average. The size of the system is chosen to be $L=100-200$,
such that all finite size effects are negligible on the simulated
time scales, and the averages are obtained using at least 100 realizations.

The integration time-step is chosen such that no qualitative influence
on the MSD is observed. For the models studied in this work time-steps
of $0.05-0.2$ were found to satisfy this criterion. Because of the
nonlinearity introduced in Eq. (\ref{eq:TDVP}) due to $P_{\mathcal{M}}$,
chaos emerges on a time-scale, dubbed the Lyapunov time, which depends
both on the bond-dimension, the realization studied, and the parameters
of the system. Beyond this time, it becomes exponentially expensive
(in time) to obtain convergence of the results on the level of \textit{individual}
configurations. We note in passing, that the Lyapunov time becomes
longer for larger bond dimension \cite{Leviatan2017}.

To assess the convergence of the method, for each configuration determined
by the initial configuration of the spins and the disorder configuration,
we calculate the convergence time, $t_{*}\left(\omega\right)$ (here
$\omega$ designates the configuration). For times $t<t_{*}\left(\omega\right)$
the dynamics generated starting from a given configuration is converged
within a required accuracy (2\%) by increasing the bond dimension.
For the infinite temperature initial condition we use in this work,
the\emph{ }convergence time, $t_{*}$, is calculated by averaging
$t_{*}\left(\omega\right)$. It is crucial to consider \emph{individual}
configurations to assess the numerical convergence of the method since
averaging over realizations introduces a fortuitous cancellation of
errors, thus while $t_{*}$ demarcates a strict, well-defined convergence
metric, apparent convergence of either transport coefficients or dynamics
may occur \emph{after} this time. The averaged convergence times for
which TDVP is numerically exact are comparable to convergence times
of conventional DMRG or MPS techniques.

\begin{figure}
\includegraphics{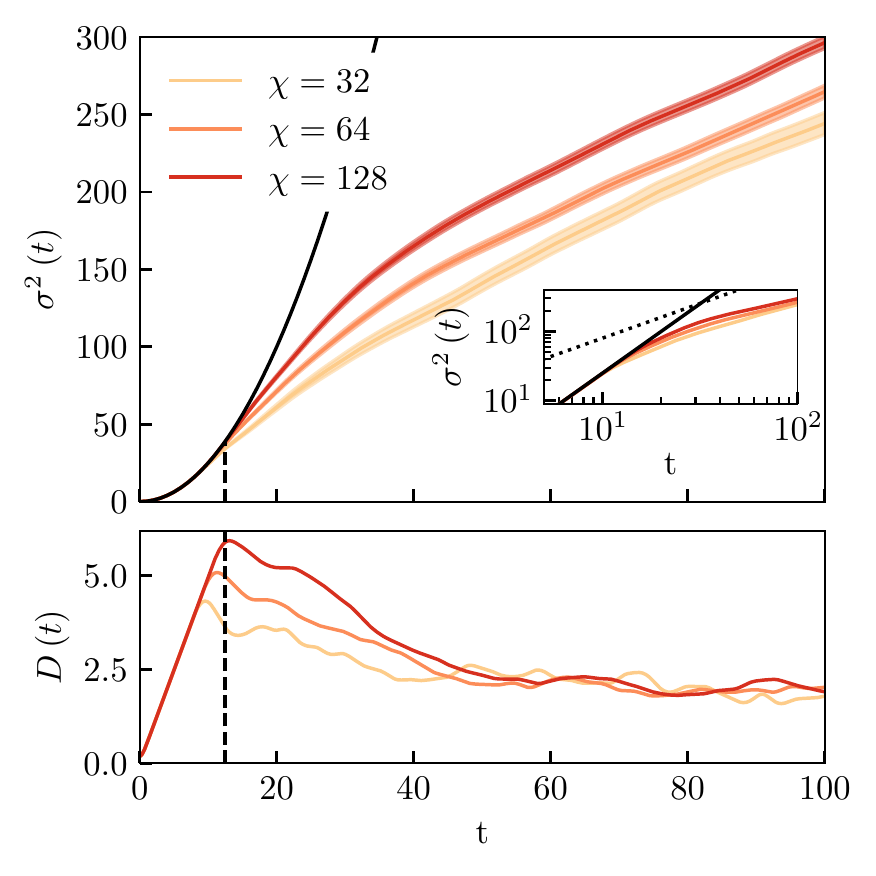}

\caption{\label{fig:XX-clean-ballistic}Clean XX model ($\Delta=0,\,W=0$).
Upper panel: MSD as a function of time for various bond dimensions
(32, 64, 128) averaged over 200-500 realizations of initial spin configurations
and disorder. More intense shades represent larger bond dimensions
and shaded areas indicate the standard-deviation of the observables
obtained using a bootstrap procedure. The black solid line is an exact
solution, obtained numerically. The inset shows the log-log scale
of the main panel with the black dotted corresponding to diffusion.
Lower panel: Time-dependent diffusion constant $D\left(t\right)$.
The dashed black line on both plots represents the convergence time,
$t_{*}$.}
\end{figure}
\begin{figure}
\includegraphics{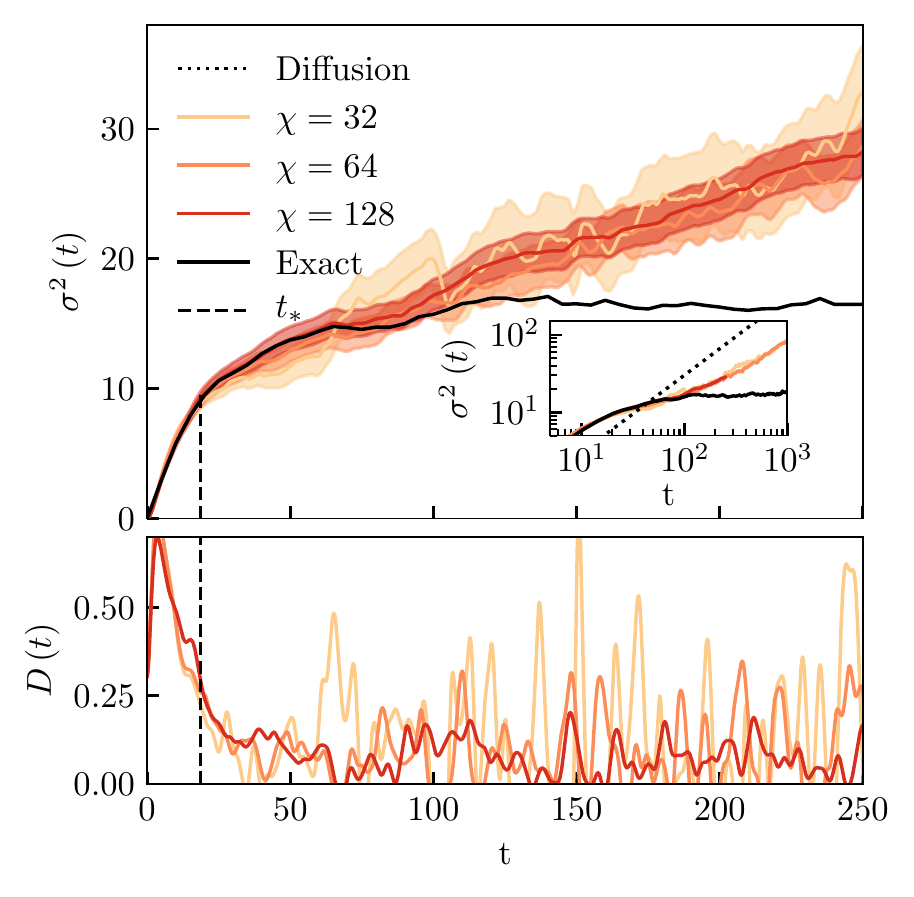}

\caption{\label{fig:MSD-XX-disordered:}Same as Fig.~(\ref{fig:XX-clean-ballistic})
but for the disordered XX model ($\Delta=0,\,W=1$) for 100 realizations
of initial spin configurations and disorder. }
\end{figure}
\begin{figure}
\includegraphics{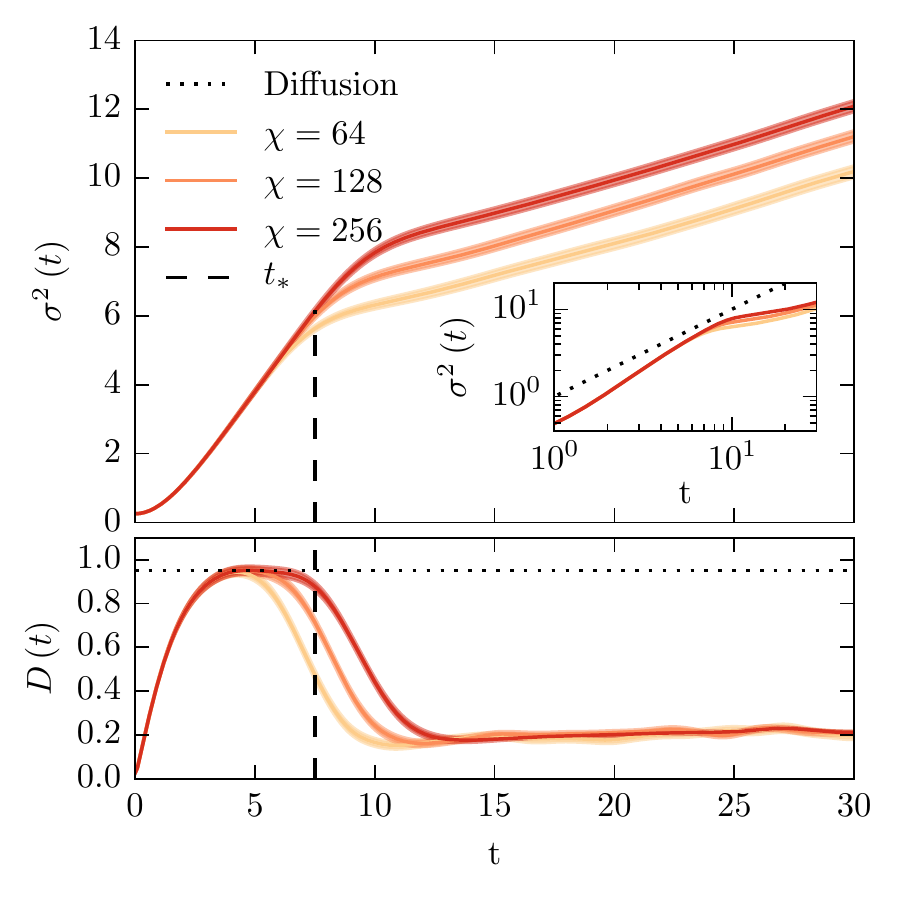}

\caption{\label{fig:XX-ladder}Same as Fig.~(\ref{fig:XX-clean-ballistic})
but for clean XX ladder of length $L=50$ with isotropic coupling
between the rungs. The results were obtained by averaging $100$ realizations
of initial spin configurations. The black dotted line in the bottom
panel represents the previously reported diffusion constant \cite{Steinigeweg2014}.}
\end{figure}
\begin{figure}
\includegraphics{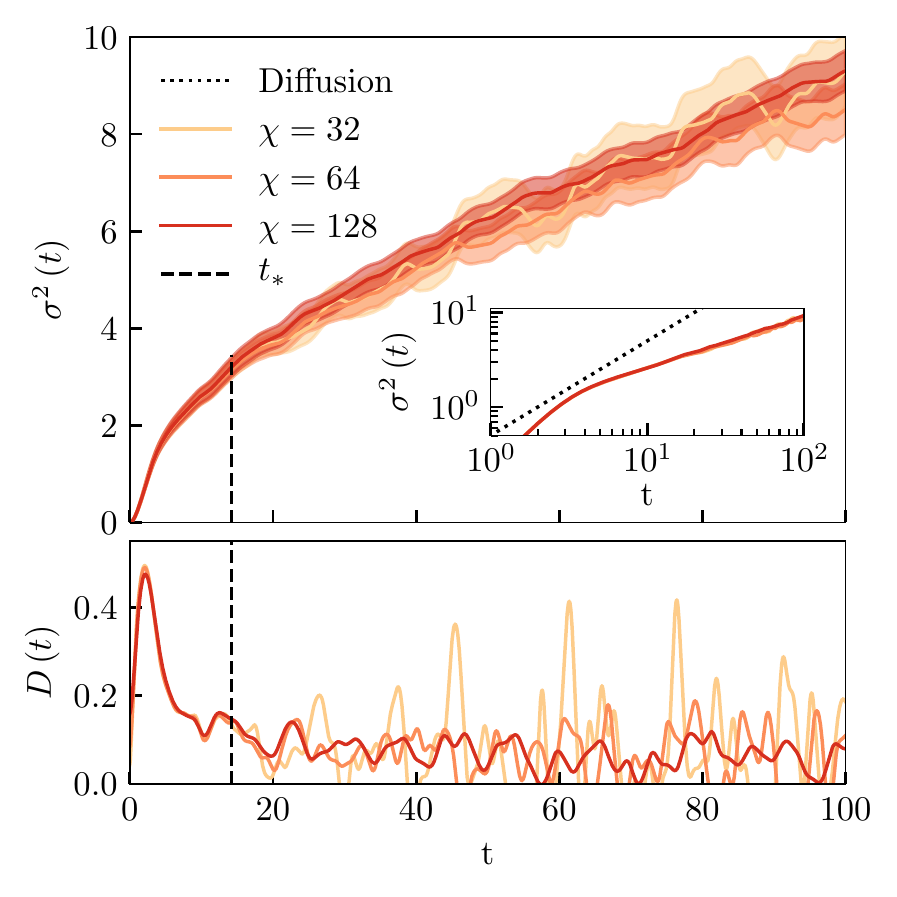}

\caption{\label{fig:MSD-XXZ-ergodic:}Same as Fig.~(\ref{fig:XX-clean-ballistic})
but for disordered XXZ model in the subdiffusive regime ($\Delta=1,\,W=1.5$)
for 200 realizations of initial spin configurations and disorder.}
\end{figure}

We first demonstrate that the long time limit essential for the study
of hydrodynamics properties and the large bond-dimension limit, when
the method becomes exact, do not ``commute,'' in the sense that
spurious, apparently converged long time behavior may emerge. For
this purpose we will first consider two integrable models with a known
dynamical behavior. We stress that true hydrodynamic behavior (at
least in the usual sense) is \emph{not} expected for such models.

\paragraph{Ballistic regime ($\Delta=0,\,W=0,\,L=200$).}

The expected ballistic transport is accurately reproduced only up
to $t_{*}\backsimeq12$ for the largest bond-dimension employed (see
Fig.~\ref{fig:XX-clean-ballistic}). While this system corresponds
to free fermions, the entanglement still grows limiting the accessible
times. Beyond the convergence time transport appears to be diffusive
with a diffusion constant of approximately $2.0$. There is little
variation of this value across the different bond-dimensions.

\paragraph{Anderson localized regime ($\Delta=0,\,W=1,\,L=150$).}

This system is also effectively noninteracting with a MSD which saturates
in time, indicating localization. TDVP fails to reproduce the plateau
for all studied bond dimensions and displays growth of the MSD with
time although the diffusion coefficient is rather small (see Fig.~\ref{fig:MSD-XX-disordered:}).
Results obtained using the largest bond-dimension (128) follow the
exact result closely up to about $t=70$, while those of smaller bond-dimensions
deviate significantly earlier, resulting in $t_{*}=19$.

Since asymptotically the nonlinear equations of TDVP are expected
to result in diffusion, the striking failure of the method for the
two integrable systems above is not surprising. 

\paragraph{Diffusive XX-ladder ($\Delta=0,\,L=50$).}

This model is a generalization of (\ref{eq:XXZ-Ham}) to a two leg
ladder. It is nonintegrable and shows convincing diffusion with a
diffusion coefficient of about $D\sim0.95$ \cite{Steinigeweg2014,Karrasch2015b}.
As expected for short times the calculations based on the TDVP reproduce
this numerically exact results (see Fig.~\ref{fig:XX-ladder}) However,
for times longer than the convergence time, $t>t_{*}=8,$ a crossover
to yet another diffusive regime with much lower diffusion constant
appears $\left(D\sim0.2\right)$. Moreover this diffusion coefficient
does not appear to strongly depend on the bond-dimension. 

The above examples illustrate that the seemingly converged transport
coefficients and long time dynamics within the TDVP framework can
be highly misleading. After demonstrating the pitfalls in determining
the long-time properties using TDVP, we examine its potential as a
hydrodynamic method for a disordered nonintegrable system. 

\paragraph{Subdiffusive regime ($\Delta=1.0,\,W=1.5,\,L=100$).}

For moderate disorder $0<W<3.7$ the system is nonintegrable \cite{Luitz2016c}.
While the convergence time here is about $t_{*}=18$, semi-quantatively
similar subdiffusive transport appears also at much longer times (see
inset in Fig.~\ref{fig:MSD-XXZ-ergodic:}). Interestingly, the exponent
extracted from the long-time behavior, $\alpha=0.54$, is in excellent
agreement with previously reported value, extracted from the short
time dynamics of the same system using exact diagonalization \cite{Lev2014,Agarwal2014,Luitz2015a,Luitz2016c}.
This indicates that for such a system, true asymptotic dynamical behavior
may indeed be uncovered using moderate numerical costs (small bond-dimensions).

\emph{Discussion}.\textemdash{} In this work we have examined how
well TDVP captures the \emph{long time} behavior of quantum interacting
systems. For any \emph{finite} time the method is formally numerically
exact, since it can be converged with respect to the bond dimension
and other numerical parameters. For longer times convergence cannot
be guaranteed generically, but one hopes that \emph{on average} the
method will still produce the correct result, due to ergodicity of
the TDVP trajectories on the MPS manifold. This assumes that the MPS
ansatz captures all the relevant local correlations that produce long
time behavior.

By examining integrable and nonintegrable models for which the asymptotic
dynamics in known, we have shown that the apparent convergence of
long time observables, such as the diffusion coefficient, obtained
using TDVP is \emph{not} indicative of the accuracy of the method
and may be very misleading. While the dramatic failure of TDVP to
reproduce ballistic and localized dynamics is expected, it is quite
unfortunate that the method appears to fail also for a nonintegrable
diffusive model. 

Interestingly, the most promising results are obtained for the nonintegrable
\emph{disordered} XXZ model in the ergodic subdiffusive phase \cite{Luitz2016c},
which is the only presented example where the short time and long
time behavior appear to agree very well, although the same caveats
concerning convergence apply. This is quite surprising, in light of
the expectation of asymptotic diffusion in TDVP generated dynamics
due to the underlying nonlinearity of the equations of motion. Nevertheless,
we find that the MSD calculated by TDVP is strongly sublinear, although
we cannot rule out a slow approach of the dynamical exponent to its
diffusive value. We would like to point out a possible connection
between the nonlinearity introduced by the tangent space projector
into the TDVP equations of motion and the nonlinear dependence on
the wave function in the self-consistent second Born approximation,
\cite{Basko2006a,BarLev2014,BarLev2015} and the nonlinear Schrödinger
equation (NLSE), both of which also show subdiffusive transport in
the presence of disorder \cite{Fishman2012}.

In summary, we have shown that great care must be exercised examining
the apparent convergence of long time properties within the TDVP approach,
which appears to generically produce either qualitatively or quantitatively
incorrect results. Nevertheless, we have presented one nontrivial
system were the short time (numerically exact) dynamics and the long
time dynamics agree, and therefore hint at the possibility of an accurate
asymptotic description, obtained at a modest computational effort.
It is of great importance to further investigate the origins of the
apparent success of the method in this case as well as to extend this
study to other nonintegrable systems in one and two-dimensions.
\begin{acknowledgments}
We would like to thank Fabian Heidrich-Meisner for suggesting the
use of the XX-ladder model. DRR and YB acknowledges funding from the
Simons Foundation (\#454951, David R. Reichman). BK acknowledges funding
through the Edith \& Eugene Blout Fellowship. This work used the Extreme
Science and Engineering Discovery Environment (XSEDE), which is supported
by National Science Foundation Grant No. OCI-1053575. 
\end{acknowledgments}

\bibliographystyle{apsrev4-1}
\bibliography{lib_yevgeny,library_link}

\begin{thebibliography}{27}%
\makeatletter
\providecommand \@ifxundefined [1]{%
 \@ifx{#1\undefined}
}%
\providecommand \@ifnum [1]{%
 \ifnum #1\expandafter \@firstoftwo
 \else \expandafter \@secondoftwo
 \fi
}%
\providecommand \@ifx [1]{%
 \ifx #1\expandafter \@firstoftwo
 \else \expandafter \@secondoftwo
 \fi
}%
\providecommand \natexlab [1]{#1}%
\providecommand \enquote  [1]{``#1''}%
\providecommand \bibnamefont  [1]{#1}%
\providecommand \bibfnamefont [1]{#1}%
\providecommand \citenamefont [1]{#1}%
\providecommand \href@noop [0]{\@secondoftwo}%
\providecommand \href [0]{\begingroup \@sanitize@url \@href}%
\providecommand \@href[1]{\@@startlink{#1}\@@href}%
\providecommand \@@href[1]{\endgroup#1\@@endlink}%
\providecommand \@sanitize@url [0]{\catcode `\\12\catcode `\$12\catcode
  `\&12\catcode `\#12\catcode `\^12\catcode `\_12\catcode `\%12\relax}%
\providecommand \@@startlink[1]{}%
\providecommand \@@endlink[0]{}%
\providecommand \url  [0]{\begingroup\@sanitize@url \@url }%
\providecommand \@url [1]{\endgroup\@href {#1}{\urlprefix }}%
\providecommand \urlprefix  [0]{URL }%
\providecommand \Eprint [0]{\href }%
\providecommand \doibase [0]{http://dx.doi.org/}%
\providecommand \selectlanguage [0]{\@gobble}%
\providecommand \bibinfo  [0]{\@secondoftwo}%
\providecommand \bibfield  [0]{\@secondoftwo}%
\providecommand \translation [1]{[#1]}%
\providecommand \BibitemOpen [0]{}%
\providecommand \bibitemStop [0]{}%
\providecommand \bibitemNoStop [0]{.\EOS\space}%
\providecommand \EOS [0]{\spacefactor3000\relax}%
\providecommand \BibitemShut  [1]{\csname bibitem#1\endcsname}%
\let\auto@bib@innerbib\@empty
\bibitem [{\citenamefont {White}(1992)}]{White1992}%
  \BibitemOpen
  \bibfield  {author} {\bibinfo {author} {\bibfnamefont {S.~R.}\ \bibnamefont
  {White}},\ }\href {\doibase 10.1103/PhysRevLett.69.2863} {\bibfield
  {journal} {\bibinfo  {journal} {Physical Review Letters}\ }\textbf {\bibinfo
  {volume} {69}},\ \bibinfo {pages} {2863} (\bibinfo {year}
  {1992})}\BibitemShut {NoStop}%
\bibitem [{\citenamefont {Schollw{\"{o}}ck}(2011)}]{Schollwock2011}%
  \BibitemOpen
  \bibfield  {author} {\bibinfo {author} {\bibfnamefont {U.}~\bibnamefont
  {Schollw{\"{o}}ck}},\ }\href {\doibase 10.1016/j.aop.2010.09.012} {\bibfield
  {journal} {\bibinfo  {journal} {Annals of Physics}\ }\textbf {\bibinfo
  {volume} {326}},\ \bibinfo {pages} {96} (\bibinfo {year} {2011})}\BibitemShut
  {NoStop}%
\bibitem [{\citenamefont {Dirac}(1930)}]{PSP:2040328}%
  \BibitemOpen
  \bibfield  {author} {\bibinfo {author} {\bibfnamefont {P.~A.~M.}\
  \bibnamefont {Dirac}},\ }\href {\doibase 10.1017/S0305004100016108}
  {\bibfield  {journal} {\bibinfo  {journal} {Mathematical Proceedings of the
  Cambridge Philosophical Society}\ }\textbf {\bibinfo {volume} {26}},\
  \bibinfo {pages} {376} (\bibinfo {year} {1930})}\BibitemShut {NoStop}%
\bibitem [{\citenamefont {Frenkel}(1934)}]{frenkel1934mechanics}%
  \BibitemOpen
  \bibfield  {author} {\bibinfo {author} {\bibfnamefont {J.~I.}\ \bibnamefont
  {Frenkel}},\ }\href@noop {} {\emph {\bibinfo {title} {{Wave mechanics :
  advanced general theory}}}},\ The international series of monographs on
  physics\ (\bibinfo  {publisher} {Clarendon Press},\ \bibinfo {address}
  {Oxford},\ \bibinfo {year} {1934})\BibitemShut {NoStop}%
\bibitem [{\citenamefont {Haegeman}\ \emph {et~al.}(2011)\citenamefont
  {Haegeman}, \citenamefont {Cirac}, \citenamefont {Osborne}, \citenamefont
  {Pi{\v{z}}orn}, \citenamefont {Verschelde},\ and\ \citenamefont
  {Verstraete}}]{Haegeman2011}%
  \BibitemOpen
  \bibfield  {author} {\bibinfo {author} {\bibfnamefont {J.}~\bibnamefont
  {Haegeman}}, \bibinfo {author} {\bibfnamefont {J.~I.}\ \bibnamefont {Cirac}},
  \bibinfo {author} {\bibfnamefont {T.~J.}\ \bibnamefont {Osborne}}, \bibinfo
  {author} {\bibfnamefont {I.}~\bibnamefont {Pi{\v{z}}orn}}, \bibinfo {author}
  {\bibfnamefont {H.}~\bibnamefont {Verschelde}}, \ and\ \bibinfo {author}
  {\bibfnamefont {F.}~\bibnamefont {Verstraete}},\ }\href {\doibase
  10.1103/PhysRevLett.107.070601} {\bibfield  {journal} {\bibinfo  {journal}
  {Physical Review Letters}\ }\textbf {\bibinfo {volume} {107}},\ \bibinfo
  {pages} {070601} (\bibinfo {year} {2011})}\BibitemShut {NoStop}%
\bibitem [{\citenamefont {Haegeman}\ \emph {et~al.}(2016)\citenamefont
  {Haegeman}, \citenamefont {Lubich}, \citenamefont {Oseledets}, \citenamefont
  {Vandereycken},\ and\ \citenamefont {Verstraete}}]{Haegeman2016}%
  \BibitemOpen
  \bibfield  {author} {\bibinfo {author} {\bibfnamefont {J.}~\bibnamefont
  {Haegeman}}, \bibinfo {author} {\bibfnamefont {C.}~\bibnamefont {Lubich}},
  \bibinfo {author} {\bibfnamefont {I.}~\bibnamefont {Oseledets}}, \bibinfo
  {author} {\bibfnamefont {B.}~\bibnamefont {Vandereycken}}, \ and\ \bibinfo
  {author} {\bibfnamefont {F.}~\bibnamefont {Verstraete}},\ }\href {\doibase
  10.1103/PhysRevB.94.165116} {\bibfield  {journal} {\bibinfo  {journal}
  {Physical Review B}\ }\textbf {\bibinfo {volume} {94}},\ \bibinfo {pages}
  {165116} (\bibinfo {year} {2016})}\BibitemShut {NoStop}%
\bibitem [{\citenamefont {Haegeman}\ \emph {et~al.}(2013)\citenamefont
  {Haegeman}, \citenamefont {Osborne},\ and\ \citenamefont
  {Verstraete}}]{Haegeman2013}%
  \BibitemOpen
  \bibfield  {author} {\bibinfo {author} {\bibfnamefont {J.}~\bibnamefont
  {Haegeman}}, \bibinfo {author} {\bibfnamefont {T.~J.}\ \bibnamefont
  {Osborne}}, \ and\ \bibinfo {author} {\bibfnamefont {F.}~\bibnamefont
  {Verstraete}},\ }\href {\doibase 10.1103/PhysRevB.88.075133} {\bibfield
  {journal} {\bibinfo  {journal} {Physical Review B}\ }\textbf {\bibinfo
  {volume} {88}},\ \bibinfo {pages} {075133} (\bibinfo {year}
  {2013})}\BibitemShut {NoStop}%
\bibitem [{\citenamefont {Leviatan}\ \emph {et~al.}(2017)\citenamefont
  {Leviatan}, \citenamefont {Pollmann}, \citenamefont {Bardarson},\ and\
  \citenamefont {Altman}}]{Leviatan2017}%
  \BibitemOpen
  \bibfield  {author} {\bibinfo {author} {\bibfnamefont {E.}~\bibnamefont
  {Leviatan}}, \bibinfo {author} {\bibfnamefont {F.}~\bibnamefont {Pollmann}},
  \bibinfo {author} {\bibfnamefont {J.~H.}\ \bibnamefont {Bardarson}}, \ and\
  \bibinfo {author} {\bibfnamefont {E.}~\bibnamefont {Altman}},\ }\href
  {http://arxiv.org/abs/1702.08894} {\enquote {\bibinfo {title} {{Quantum
  thermalization dynamics with Matrix-Product States}},}\ } (\bibinfo {year}
  {2017}),\ \Eprint {http://arxiv.org/abs/1702.08894} {arXiv:1702.08894}
  \BibitemShut {NoStop}%
\bibitem [{\citenamefont {Bertini}\ \emph {et~al.}(2016)\citenamefont
  {Bertini}, \citenamefont {Collura}, \citenamefont {De~Nardis},\ and\
  \citenamefont {Fagotti}}]{PhysRevLett.117.207201}%
  \BibitemOpen
  \bibfield  {author} {\bibinfo {author} {\bibfnamefont {B.}~\bibnamefont
  {Bertini}}, \bibinfo {author} {\bibfnamefont {M.}~\bibnamefont {Collura}},
  \bibinfo {author} {\bibfnamefont {J.}~\bibnamefont {De~Nardis}}, \ and\
  \bibinfo {author} {\bibfnamefont {M.}~\bibnamefont {Fagotti}},\ }\href
  {\doibase 10.1103/PhysRevLett.117.207201} {\bibfield  {journal} {\bibinfo
  {journal} {Phys. Rev. Lett.}\ }\textbf {\bibinfo {volume} {117}},\ \bibinfo
  {pages} {207201} (\bibinfo {year} {2016})}\BibitemShut {NoStop}%
\bibitem [{\citenamefont {Jordan}\ and\ \citenamefont
  {Wigner}(1928)}]{Jordan1928}%
  \BibitemOpen
  \bibfield  {author} {\bibinfo {author} {\bibfnamefont {P.}~\bibnamefont
  {Jordan}}\ and\ \bibinfo {author} {\bibfnamefont {E.}~\bibnamefont
  {Wigner}},\ }\href {\doibase 10.1007/BF01331938} {\bibfield  {journal}
  {\bibinfo  {journal} {Zeitschrift f{\"{u}}r Physik}\ }\textbf {\bibinfo
  {volume} {47}},\ \bibinfo {pages} {631} (\bibinfo {year} {1928})}\BibitemShut
  {NoStop}%
\bibitem [{\citenamefont {Anderson}(1958)}]{Anderson1958b}%
  \BibitemOpen
  \bibfield  {author} {\bibinfo {author} {\bibfnamefont {P.~W.}\ \bibnamefont
  {Anderson}},\ }\href {\doibase 10.1103/PhysRev.109.1492} {\bibfield
  {journal} {\bibinfo  {journal} {Physical Review}\ }\textbf {\bibinfo {volume}
  {109}},\ \bibinfo {pages} {1492} (\bibinfo {year} {1958})}\BibitemShut
  {NoStop}%
\bibitem [{\citenamefont {Basko}\ \emph {et~al.}(2006)\citenamefont {Basko},
  \citenamefont {Aleiner},\ and\ \citenamefont {Altshuler}}]{Basko2006a}%
  \BibitemOpen
  \bibfield  {author} {\bibinfo {author} {\bibfnamefont {D.}~\bibnamefont
  {Basko}}, \bibinfo {author} {\bibfnamefont {I.~L.}\ \bibnamefont {Aleiner}},
  \ and\ \bibinfo {author} {\bibfnamefont {B.~L.}\ \bibnamefont {Altshuler}},\
  }\href {\doibase 10.1016/j.aop.2005.11.014} {\bibfield  {journal} {\bibinfo
  {journal} {Annals of Physics}\ }\textbf {\bibinfo {volume} {321}},\ \bibinfo
  {pages} {1126} (\bibinfo {year} {2006})}\BibitemShut {NoStop}%
\bibitem [{\citenamefont {Abanin}\ and\ \citenamefont
  {Papi{\'{c}}}(2017)}]{Abanin2017}%
  \BibitemOpen
  \bibfield  {author} {\bibinfo {author} {\bibfnamefont {D.~A.}\ \bibnamefont
  {Abanin}}\ and\ \bibinfo {author} {\bibfnamefont {Z.}~\bibnamefont
  {Papi{\'{c}}}},\ }\href {http://arxiv.org/abs/1705.09103} {\enquote {\bibinfo
  {title} {{Recent progress in many-body localization}},}\ } (\bibinfo {year}
  {2017}),\ \Eprint {http://arxiv.org/abs/1705.09103} {arXiv:1705.09103}
  \BibitemShut {NoStop}%
\bibitem [{\citenamefont {Berkelbach}\ and\ \citenamefont
  {Reichman}(2010)}]{Berkelbach2010a}%
  \BibitemOpen
  \bibfield  {author} {\bibinfo {author} {\bibfnamefont {T.~C.}\ \bibnamefont
  {Berkelbach}}\ and\ \bibinfo {author} {\bibfnamefont {D.~R.}\ \bibnamefont
  {Reichman}},\ }\href {\doibase 10.1103/PhysRevB.81.224429} {\bibfield
  {journal} {\bibinfo  {journal} {Physical Review B}\ }\textbf {\bibinfo
  {volume} {81}},\ \bibinfo {pages} {224429} (\bibinfo {year}
  {2010})}\BibitemShut {NoStop}%
\bibitem [{\citenamefont {Luitz}\ \emph {et~al.}(2015)\citenamefont {Luitz},
  \citenamefont {Laflorencie},\ and\ \citenamefont {Alet}}]{Luitz2015}%
  \BibitemOpen
  \bibfield  {author} {\bibinfo {author} {\bibfnamefont {D.~J.}\ \bibnamefont
  {Luitz}}, \bibinfo {author} {\bibfnamefont {N.}~\bibnamefont {Laflorencie}},
  \ and\ \bibinfo {author} {\bibfnamefont {F.}~\bibnamefont {Alet}},\ }\href
  {\doibase 10.1103/PhysRevB.91.081103} {\bibfield  {journal} {\bibinfo
  {journal} {Physical Review B}\ }\textbf {\bibinfo {volume} {91}},\ \bibinfo
  {pages} {081103} (\bibinfo {year} {2015})}\BibitemShut {NoStop}%
\bibitem [{\citenamefont {Steinigeweg}\ \emph {et~al.}(2009)\citenamefont
  {Steinigeweg}, \citenamefont {Wichterich},\ and\ \citenamefont
  {Gemmer}}]{Steinigeweg2009a}%
  \BibitemOpen
  \bibfield  {author} {\bibinfo {author} {\bibfnamefont {R.}~\bibnamefont
  {Steinigeweg}}, \bibinfo {author} {\bibfnamefont {H.}~\bibnamefont
  {Wichterich}}, \ and\ \bibinfo {author} {\bibfnamefont {J.}~\bibnamefont
  {Gemmer}},\ }\href {\doibase 10.1209/0295-5075/88/10004} {\bibfield
  {journal} {\bibinfo  {journal} {EPL (Europhysics Letters)}\ }\textbf
  {\bibinfo {volume} {88}},\ \bibinfo {pages} {10004} (\bibinfo {year}
  {2009})}\BibitemShut {NoStop}%
\bibitem [{\citenamefont {Yan}\ \emph {et~al.}(2015)\citenamefont {Yan},
  \citenamefont {Jiang},\ and\ \citenamefont {Zhao}}]{Yan2015}%
  \BibitemOpen
  \bibfield  {author} {\bibinfo {author} {\bibfnamefont {Y.}~\bibnamefont
  {Yan}}, \bibinfo {author} {\bibfnamefont {F.}~\bibnamefont {Jiang}}, \ and\
  \bibinfo {author} {\bibfnamefont {H.}~\bibnamefont {Zhao}},\ }\href {\doibase
  10.1140/epjb/e2014-50797-4} {\bibfield  {journal} {\bibinfo  {journal} {The
  European Physical Journal B}\ }\textbf {\bibinfo {volume} {88}},\ \bibinfo
  {pages} {11} (\bibinfo {year} {2015})}\BibitemShut {NoStop}%
\bibitem [{\citenamefont {Steinigeweg}\ \emph {et~al.}(2017)\citenamefont
  {Steinigeweg}, \citenamefont {Jin}, \citenamefont {Schmidtke}, \citenamefont
  {{De Raedt}}, \citenamefont {Michielsen},\ and\ \citenamefont
  {Gemmer}}]{Steinigeweg2017}%
  \BibitemOpen
  \bibfield  {author} {\bibinfo {author} {\bibfnamefont {R.}~\bibnamefont
  {Steinigeweg}}, \bibinfo {author} {\bibfnamefont {F.}~\bibnamefont {Jin}},
  \bibinfo {author} {\bibfnamefont {D.}~\bibnamefont {Schmidtke}}, \bibinfo
  {author} {\bibfnamefont {H.}~\bibnamefont {{De Raedt}}}, \bibinfo {author}
  {\bibfnamefont {K.}~\bibnamefont {Michielsen}}, \ and\ \bibinfo {author}
  {\bibfnamefont {J.}~\bibnamefont {Gemmer}},\ }\href {\doibase
  10.1103/PhysRevB.95.035155} {\bibfield  {journal} {\bibinfo  {journal}
  {Physical Review B}\ }\textbf {\bibinfo {volume} {95}},\ \bibinfo {pages}
  {035155} (\bibinfo {year} {2017})}\BibitemShut {NoStop}%
\bibitem [{\citenamefont {Luitz}\ and\ \citenamefont {{Bar
  Lev}}(2017)}]{Luitz2016c}%
  \BibitemOpen
  \bibfield  {author} {\bibinfo {author} {\bibfnamefont {D.~J.}\ \bibnamefont
  {Luitz}}\ and\ \bibinfo {author} {\bibfnamefont {Y.}~\bibnamefont {{Bar
  Lev}}},\ }\href {\doibase 10.1002/andp.201600350} {\bibfield  {journal}
  {\bibinfo  {journal} {Annalen der Physik}\ }\textbf {\bibinfo {volume}
  {529}},\ \bibinfo {pages} {1600350} (\bibinfo {year} {2017})}\BibitemShut
  {NoStop}%
\bibitem [{\citenamefont {Steinigeweg}\ \emph {et~al.}(2014)\citenamefont
  {Steinigeweg}, \citenamefont {Heidrich-Meisner}, \citenamefont {Gemmer},
  \citenamefont {Michielsen},\ and\ \citenamefont {{De
  Raedt}}}]{Steinigeweg2014}%
  \BibitemOpen
  \bibfield  {author} {\bibinfo {author} {\bibfnamefont {R.}~\bibnamefont
  {Steinigeweg}}, \bibinfo {author} {\bibfnamefont {F.}~\bibnamefont
  {Heidrich-Meisner}}, \bibinfo {author} {\bibfnamefont {J.}~\bibnamefont
  {Gemmer}}, \bibinfo {author} {\bibfnamefont {K.}~\bibnamefont {Michielsen}},
  \ and\ \bibinfo {author} {\bibfnamefont {H.}~\bibnamefont {{De Raedt}}},\
  }\href {\doibase 10.1103/PhysRevB.90.094417} {\bibfield  {journal} {\bibinfo
  {journal} {Physical Review B}\ }\textbf {\bibinfo {volume} {90}},\ \bibinfo
  {pages} {094417} (\bibinfo {year} {2014})}\BibitemShut {NoStop}%
\bibitem [{\citenamefont {Karrasch}\ \emph {et~al.}(2015)\citenamefont
  {Karrasch}, \citenamefont {Kennes},\ and\ \citenamefont
  {Heidrich-Meisner}}]{Karrasch2015b}%
  \BibitemOpen
  \bibfield  {author} {\bibinfo {author} {\bibfnamefont {C.}~\bibnamefont
  {Karrasch}}, \bibinfo {author} {\bibfnamefont {D.~M.}\ \bibnamefont
  {Kennes}}, \ and\ \bibinfo {author} {\bibfnamefont {F.}~\bibnamefont
  {Heidrich-Meisner}},\ }\href {\doibase 10.1103/PhysRevB.91.115130} {\bibfield
   {journal} {\bibinfo  {journal} {Physical Review B}\ }\textbf {\bibinfo
  {volume} {91}},\ \bibinfo {pages} {115130} (\bibinfo {year}
  {2015})}\BibitemShut {NoStop}%
\bibitem [{\citenamefont {{Bar Lev}}\ \emph {et~al.}(2015)\citenamefont {{Bar
  Lev}}, \citenamefont {Cohen},\ and\ \citenamefont {Reichman}}]{Lev2014}%
  \BibitemOpen
  \bibfield  {author} {\bibinfo {author} {\bibfnamefont {Y.}~\bibnamefont {{Bar
  Lev}}}, \bibinfo {author} {\bibfnamefont {G.}~\bibnamefont {Cohen}}, \ and\
  \bibinfo {author} {\bibfnamefont {D.~R.}\ \bibnamefont {Reichman}},\ }\href
  {\doibase 10.1103/PhysRevLett.114.100601} {\bibfield  {journal} {\bibinfo
  {journal} {Physical Review Letters}\ }\textbf {\bibinfo {volume} {114}},\
  \bibinfo {pages} {100601} (\bibinfo {year} {2015})}\BibitemShut {NoStop}%
\bibitem [{\citenamefont {Agarwal}\ \emph {et~al.}(2015)\citenamefont
  {Agarwal}, \citenamefont {Gopalakrishnan}, \citenamefont {Knap},
  \citenamefont {M{\"{u}}ller},\ and\ \citenamefont {Demler}}]{Agarwal2014}%
  \BibitemOpen
  \bibfield  {author} {\bibinfo {author} {\bibfnamefont {K.}~\bibnamefont
  {Agarwal}}, \bibinfo {author} {\bibfnamefont {S.}~\bibnamefont
  {Gopalakrishnan}}, \bibinfo {author} {\bibfnamefont {M.}~\bibnamefont
  {Knap}}, \bibinfo {author} {\bibfnamefont {M.}~\bibnamefont {M{\"{u}}ller}},
  \ and\ \bibinfo {author} {\bibfnamefont {E.}~\bibnamefont {Demler}},\ }\href
  {\doibase 10.1103/PhysRevLett.114.160401} {\bibfield  {journal} {\bibinfo
  {journal} {Physical Review Letters}\ }\textbf {\bibinfo {volume} {114}},\
  \bibinfo {pages} {160401} (\bibinfo {year} {2015})}\BibitemShut {NoStop}%
\bibitem [{\citenamefont {Luitz}\ \emph {et~al.}(2016)\citenamefont {Luitz},
  \citenamefont {Laflorencie},\ and\ \citenamefont {Alet}}]{Luitz2015a}%
  \BibitemOpen
  \bibfield  {author} {\bibinfo {author} {\bibfnamefont {D.~J.}\ \bibnamefont
  {Luitz}}, \bibinfo {author} {\bibfnamefont {N.}~\bibnamefont {Laflorencie}},
  \ and\ \bibinfo {author} {\bibfnamefont {F.}~\bibnamefont {Alet}},\ }\href
  {\doibase 10.1103/PhysRevB.93.060201} {\bibfield  {journal} {\bibinfo
  {journal} {Physical Review B}\ }\textbf {\bibinfo {volume} {93}},\ \bibinfo
  {pages} {060201} (\bibinfo {year} {2016})}\BibitemShut {NoStop}%
\bibitem [{\citenamefont {{Bar Lev}}\ and\ \citenamefont
  {Reichman}(2014)}]{BarLev2014}%
  \BibitemOpen
  \bibfield  {author} {\bibinfo {author} {\bibfnamefont {Y.}~\bibnamefont {{Bar
  Lev}}}\ and\ \bibinfo {author} {\bibfnamefont {D.~R.}\ \bibnamefont
  {Reichman}},\ }\href {\doibase 10.1103/PhysRevB.89.220201} {\bibfield
  {journal} {\bibinfo  {journal} {Physical Review B}\ }\textbf {\bibinfo
  {volume} {89}},\ \bibinfo {pages} {220201} (\bibinfo {year}
  {2014})}\BibitemShut {NoStop}%
\bibitem [{\citenamefont {{Bar Lev}}\ and\ \citenamefont
  {Reichman}(2016)}]{BarLev2015}%
  \BibitemOpen
  \bibfield  {author} {\bibinfo {author} {\bibfnamefont {Y.}~\bibnamefont {{Bar
  Lev}}}\ and\ \bibinfo {author} {\bibfnamefont {D.~R.}\ \bibnamefont
  {Reichman}},\ }\href {\doibase 10.1209/0295-5075/113/46001} {\bibfield
  {journal} {\bibinfo  {journal} {EPL (Europhysics Letters)}\ }\textbf
  {\bibinfo {volume} {113}},\ \bibinfo {pages} {46001} (\bibinfo {year}
  {2016})}\BibitemShut {NoStop}%
\bibitem [{\citenamefont {Fishman}\ \emph {et~al.}(2012)\citenamefont
  {Fishman}, \citenamefont {Krivolapov},\ and\ \citenamefont
  {Soffer}}]{Fishman2012}%
  \BibitemOpen
  \bibfield  {author} {\bibinfo {author} {\bibfnamefont {S.}~\bibnamefont
  {Fishman}}, \bibinfo {author} {\bibfnamefont {Y.}~\bibnamefont {Krivolapov}},
  \ and\ \bibinfo {author} {\bibfnamefont {A.}~\bibnamefont {Soffer}},\ }\href
  {\doibase 10.1088/0951-7715/25/4/R53} {\bibfield  {journal} {\bibinfo
  {journal} {Nonlinearity}\ }\textbf {\bibinfo {volume} {25}},\ \bibinfo
  {pages} {R53} (\bibinfo {year} {2012})}\BibitemShut {NoStop}%
\end{thebibliography}%

\end{document}